\documentclass[12pt]{article}
\usepackage{epsfig}\parskip 5pt plus 1pt
\newcommand{\be}{\begin{equation}}
\newcommand{\ee}{\end{equation}}
\newcommand{\bea}{\begin{eqnarray}}
\newcommand{\eea}{\end{eqnarray}}

\def\simge{\mathrel{%
   \rlap{\raise 0.511ex \hbox{$>$}}{\lower 0.511ex \hbox{$\sim$}}}}
\def\simle{\mathrel{
   \rlap{\raise 0.511ex \hbox{$<$}}{\lower 0.511ex \hbox{$\sim$}}}}
\newcommand{\helio}{^6{\rm He}}
\newcommand{\neon}{^{18}{\rm Ne}}

\begin{document}
\thispagestyle{empty}
%\begin{flushright}
%{\tt hep-ph/ }\\
%{Preprint LNF}
%\end{flushright}
\vspace*{1cm}
\begin{center}
{\Large{\bf Neutrino hierarchy from CP-blind
observables with high density magnetized detectors} }\\
\vspace{.5cm} 

A.~Donini$^{\rm a}$, E.~Fernandez-Martinez$^{\rm a}$,
P.~Migliozzi$^{\rm b}$, S.~Rigolin$^{\rm a}$, L.~Scotto Lavina$^{\rm
b,c}$, M.~Selvi$^{\rm d}$, T.~Tabarelli de Fatis$^{\rm e}$,
F.~Terranova$^{\rm f}$ \\
\vspace*{1cm}
$^{\rm a}$ I.F.T. and Dep. F\'{\i}sica Te\'{o}rica, U.A.M., Madrid, Spain \\
$^{\rm b}$ I.N.F.N., Sez. di Napoli, Napoli, Italy \\
$^{\rm c}$ Dip. di Fisica, Universit\`{a} ``Federico II'', Napoli, Italy \\
$^{\rm d}$ I.N.F.N., Sez. di Bologna, Bologna, Italy \\
$^{\rm e}$  Universit\`{a} di Milano Bicocca and I.N.F.N., Milano,
Italy \\
$^{\rm f}$ I.N.F.N., Laboratori Nazionali di Frascati,
Frascati (Rome), Italy \\

\end{center}

\vspace{.3cm}
\begin{abstract}
High density magnetized detectors are well suited to exploit the
outstanding purity and intensities of novel neutrino sources like 
Neutrino Factories and  Beta Beams. They can also provide
independent measurements of leptonic mixing parameters through the
observation of atmospheric muon-neutrinos. In this paper, we discuss
the combination of these observables from a multi-kton iron detector
and a high energy Beta Beam; in particular, we demonstrate that even
with moderate detector granularities the neutrino mass hierarchy can
be determined for $\theta_{13}$ values greater than 4$^\circ$.
\noindent
\end{abstract}

\vspace*{\stretch{2}}
\begin{flushleft}
% insert here the PACS number 
\end{flushleft}
%\begin{center}%

\newpage

\section{Introduction}
\label{introduction}
In the standard interpretation
of the experimental evidence for neutrino
oscillation~\cite{pdg,evidence}, the squared-mass differences among
the $\nu$ mass eigenstates ($m_1$, $m_2$ and $m_3$) are rather
hierarchical: $\Delta m^2_{21} \equiv m_2^2 - m^2_1 \simeq 1/30 \times
|\Delta m^2_{32}| \simeq |\Delta m^2_{31}|$. The small size of $\Delta
m^2_{21}$ is implied by solar and long-baseline reactor neutrino data.
In this framework, particularly enlightening are the oscillations at
the ``atmospheric scale'', i.e.  when $|\Delta m^2_{32}| L/4E \simeq
\pi/2$, L and E being the neutrino energy and path-length,
respectively. A muon-neutrino oscillating at the atmospheric scale
undergoes mainly $\nu_\mu \rightarrow \nu_\tau$ transitions. This
dominant mode is implied by long-baseline accelerator experiments,
atmospherics and short-baseline reactor data; moreover, it is
currently under test in a direct manner at CNGS~\cite{opera_NJP}. We
have no evidence of subdominant $\nu_\mu \rightarrow \nu_e$
transitions at such scale, implying a small mixing angle between the
first and third mass eigenstate ($\theta_{13}<10^\circ$ at
90\%~C.L.~\cite{reactor}). If $\theta_{13}$ is non-zero, the
subdominant $\nu_\mu \rightarrow \nu_e$ amplitude and its T or CP
conjugates encode a wealth of information. In particular, it allows
the determination of $\theta_{13}$ and the Dirac complex phase of the
leptonic mixing matrix. The $\nu_\mu \rightarrow \nu_e$ transition
probability is also perturbed by matter effects if the path of the
neutrinos through the Earth is sufficiently large. The perturbation
depends on the sign of $\Delta m^2_{31}$ and therefore it allows the
determination of the hierarchy among the neutrino masses.  These
considerations explain the enormous interest toward novel neutrino
sources operating at the atmospheric scale~\cite{review} like, for
instance, Neutrino Factories~\cite{nufact} or Beta
Beams~\cite{betabeam}. Combination of these facilities with more
traditional measurements of atmospheric neutrinos has also been
considered~\cite{waterch,Huber:2005ep,campagne}. Such combination is
particularly natural when the detector at the far location is dense
and capable of charge reconstruction: this is the case of magnetized
iron calorimeters, which can perform detailed measurements of
atmospheric $\nu_\mu$ fluxes and are recognized as ideally suited to
exploit a possible Neutrino Factory or high energy Beta
Beam~\cite{highebb,Donini:2006tt,highebb2,agarwalla,litium}. However, it is
generally believed that the contribution of atmospheric neutrinos is
marginal for any realistic configuration, unless the value of the
$\theta_{13}$ angle turns out to be very close to current
bounds~\cite{Petcov:2005rv}.  In this paper, we carry out a more
detailed analysis of the atmospheric data that can be collected by a
40-kton magnetized iron calorimeter (Sec.\ref{sec:setup}).  This
analysis is tuned to identify the occurrence of resonant transitions
in the earth and it is combined with the data that can be collected by
the same detector exposed to a high energy Beta
Beam~\cite{Donini:2006tt} at a baseline of $\sim$700 km
(Sec.\ref{sec:analysis}). The sensitivity of the atmospheric data to
the neutrino mass hierarchy does not depend on the value of $\delta$
(``CP-blind'') but it is highly deteriorated by the lack of knowledge
of $\theta_{13}$; on the other hand, the Beta Beam data exhibits a
very strong dependence on $\delta$ but provide tight constraints on
the size of the $\theta_{13}$. Their combination
(Sec.~\ref{sec:sensitivity}) results in a significantly improved
capability to determine the hierarchy of the neutrino mass
eigenstates.

\section{Atmospheric and long-baseline oscillations}
\label{sec:prob}

Current experimental data are unable to fix uniquely the hierarchy of
neutrino masses~\cite{pdg}, i.e. whether the $m_1$ eigenstate is
lighter than $m_3$ ($m_1<m_2<m_3$: ``normal hierarchy'') or heavier
($m_3<m_1<m_2$: ``inverted hierarchy''). Clearly, this pattern is of
great theoretical relevance because it allows discrimination among
models that explain neutrino masses~\cite{chen}.
In vacuum, observables sensitive to the mass pattern exist but their
exploitation is extremely challenging from the experimental point of
view~\cite{disappearance}.  However, propagation in matter can enhance
the perturbations to the transition probabilities due to the sign of
$\Delta m^2_{31}$ and, for sizable values of $\theta_{13}$ such
perturbations become observable even with present technologies. In
particular, the $\nu_e \rightarrow \nu_\mu$ transition amplitude (or
its T and CP-conjugate) encodes an explicit dependence on the sign of
$\Delta m^2_{31}$~\cite{cervera,freund}:

\begin{eqnarray}
P_{\nu_e \rightarrow \nu_\mu} & \simeq & \sin^2 2\theta_{13} \, \sin^2
\theta_{23} \frac{\sin^2[(1- \hat{A}){\Delta}]}{(1-\hat{A})^2}
\nonumber \\ & + & \alpha \sin 2\theta_{13} \, \xi \sin \delta
\sin({\Delta}) \frac{\sin(\hat{A}{\Delta})}{\hat{A}}
\frac{\sin[(1-\hat{A}){\Delta}]}{(1-\hat{A})} \nonumber \\ &+& \alpha
\sin 2\theta_{13} \, \xi \cos \delta \cos({\Delta})
\frac{\sin(\hat{A}{\Delta})}{\hat{A}} \frac{\sin[(1-\hat{A}){\Delta}]}
{(1-\hat{A})} \nonumber \\ &+& \alpha^2 \, \cos^2 \theta_{23} \sin^2
2\theta_{12} \frac{\sin^2(\hat{A}{\Delta})}{\hat{A}^2} \nonumber.
\label{equ:probmatter}
\end{eqnarray}

\noindent In this formula $\Delta \equiv \Delta m_{31}^2 L/(4 E)$ and
the terms contributing to the Jarlskog invariant are split into the
small parameter $\sin 2\theta_{13}$, the ${\cal O}(1)$ term $\xi
\equiv \cos\theta_{13} \, \sin 2\theta_{12} \, \sin 2\theta_{23}$ and
the CP term $\sin \delta$; $\hat{A} \equiv 2 \sqrt{2} G_F n_e E/\Delta
m_{31}^2$ with $G_F$ the Fermi coupling constant and $n_e$ the
electron density in matter. Note that the sign of $\hat{A}$ depends on
the sign of $\Delta m_{31}^2$ which is positive (negative) for normal
(inverted) hierarchy of neutrino masses.

\begin{figure}[htb]
\centering
\epsfig{file=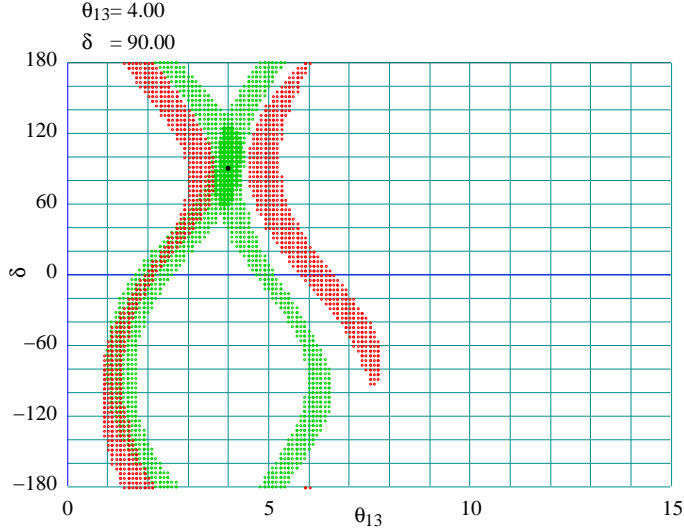,width=9cm}
\caption{Allowed $\theta_{13},\delta$ regions from a neutrino and
anti-neutrino run of a Super-SPS based Beta Beam assuming the wrong
(inverted - red) and right (normal - green) mass hierarchy
hypothesis. The true parameters are $\theta_{13}=4^\circ$,
$\mathrm{sign}(\Delta m_{31}^2)=+1$ and $\delta=90^\circ$.}
\label{fig:allowed_region_1}
\end{figure}

\begin{figure}[htb]
\centering
\epsfig{file=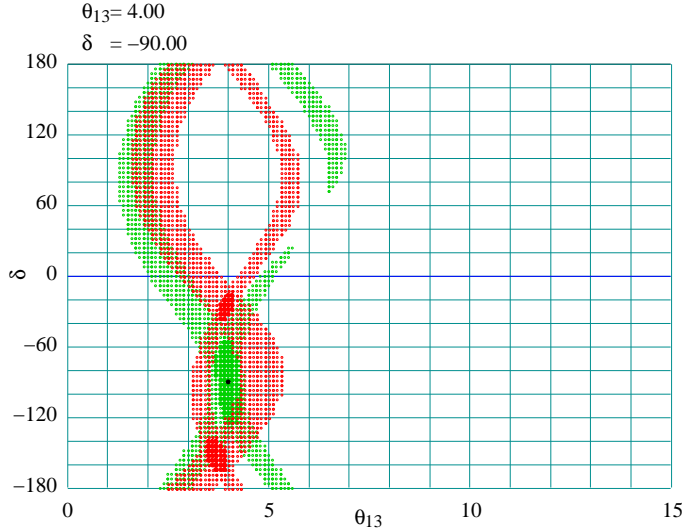,width=9cm}
\caption{As in Fig.~\ref{fig:allowed_region_1} with $\delta=-90^\circ$.
Overlapping regions are present also for the wrong hypothesis (red area).}
\label{fig:allowed_region_2}
\end{figure}

For a monochromatic beam, a $\nu_e \rightarrow \nu_\mu$ run combined
with its CP conjugate $\bar{\nu}_e \rightarrow \bar{\nu}_\mu$ cannot
determine uniquely sign($\Delta m^2_{31}$) for all values of
$\delta$. In particular, if the hierarchy is normal (inverted) and
$\delta<0$ ($\delta>0$), it is always possible to find a solution that
reproduces the correct $\nu_e \rightarrow \nu_\mu$ and $\bar{\nu}_e
\rightarrow \bar{\nu}_\mu$ rate assuming the wrong hypothesis on
sign($\Delta m^2_{31}$). For beams of finite width, spectral
information help lifting this ambiguity but, in general, they require
very large statistics and excellent detector resolution.
Figs.~\ref{fig:allowed_region_1} and \ref{fig:allowed_region_2} show
the allowed $\theta_{13},\delta$ region from a neutrino and
anti-neutrino run of a Super-SPS~\cite{supersps} based Beta Beam
assuming both the wrong (red) and right (green)
hypothesis\footnote{Details on this experimental setup are given
in~\cite{Donini:2006tt} and briefly recalled in
Sec.~\ref{sec:setup}.}. The true parameters are $\theta_{13}=4^\circ$,
sign$(\Delta m_{31}^2)=+1$ and $\delta=90^\circ$
(Fig.\ref{fig:allowed_region_1}) or $\delta=-90^\circ$
(Fig.\ref{fig:allowed_region_2}). The existence of overlapping regions
in Fig.\ref{fig:allowed_region_2} testifies the consistency of the
wrong hypothesis with the accelerator data. In general, we expect a
highly deteriorated sensitivity to mass hierarchy for nearly 50\% of
the possible true values of $\delta$. This is clearly visible in Fig.8
of \cite{Donini:2006tt} and it is reproduced in
Fig.\ref{fig:sensitivity} (blue line)\footnote{Actually, most of the
next generation long-baseline experiments proposed so far suffer from
this limitation. For an example based on Superbeams we refer
to~\cite{nova}.}.

\begin{figure}[htb]
\centering
\epsfig{file=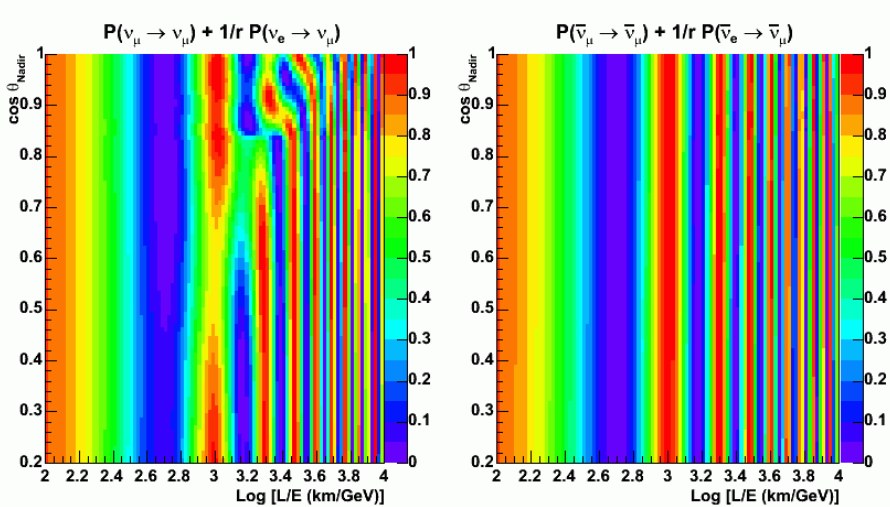,width=12cm}
\caption{Probability of observing atmospheric muon neutrinos (left
plot - antineutrinos are shown in the right plot) from $\nu_\mu
\rightarrow \nu_\mu$ and $\nu_e \rightarrow \nu_\mu$ transitions as a
function of L/E and the Nadir angle $\theta$ ($\sin^2
2\theta_{13}=0.1$, $\Delta m^2_{31} = 2.5 \cdot 10^{-3}$~eV$^2$).
In the upper label, $r$ is
the ratio between the initial $\nu_\mu$ and $\nu_e$ fluxes.}
\label{fig:proboscmu}
\end{figure}

Since the dependence on $\delta$ and, in general, the three-family
interference effects are the origin of this ambiguity, a substantial
improvement can be achieved exploiting CP-blind observables. In $\nu_e
\rightarrow \nu_\mu$ transitions, CP-blindness can be achieved
introducing a phase advance (increase of baseline) such that the phase
acquired by neutrinos due to interaction with matter equals
$2\pi$~\cite{magic}. It can be shown~\cite{Smirnov:2006sm} that this
condition is equivalent to choosing $L=2\pi/\sqrt{2}G_F n_e$,
i.e. finding a $L$ such that the last three terms of
Eq.\ref{equ:probmatter} cancels. Such baseline is dubbed ``magic'' in
literature~\cite{magic} and it is proportional to the energy and width
of the MSW resonance in a medium of constant density:
\be
E_R= \pm \Delta m^2_{31} L_{magic} \ \cos 2\theta_{13}  /4 \pi
\label{equ:energy}
\ee
\be
\Gamma_R= |\Delta m^2_{31}| L_{magic} \ \sin 2\theta_{13} /2 \pi
\label{equ:width}
\ee
the $+$ ($-$) sign referring to neutrinos (antineutrinos). A similar
approach can be pursued using atmospheric neutrinos. In this case,
however, additional difficulties are present. First of all, the
initial flux is a mixture of $\nu_\mu$, $\nu_e$ and their
antiparticles. Moreover, the approximation of constant density is non
tenable and core-mantle interference effects are possible. Finally,
the detector resolution must be sufficient to identify the region
where the MSW resonance occurs. In turn, the latter condition is a
requirement on the size of $\theta_{13}$ through Eq.\ref{equ:width}.
Eq.~\ref{equ:energy} implies that the resonance condition can occur
only for neutrinos if the hierarchy is normal and only for
anti-neutrinos if the hierarchy is inverted. Hence, the task of fixing
the hierarchy is reduced to the issue of determining the occurrence of
this resonance or at least the spectral perturbation induced by it in
the neutrino (antineutrino) sample. Such perturbation is depicted in
Fig.\ref{fig:proboscmu} for $\sin^2 2\theta_{13}=0.1$ (normal
hierarchy) as a function of L/E and the Nadir angle $\theta$ 
%(see Fig.~\ref{fig:schema}) 
of the neutrinos. It has been computed solving
numerically~\cite{tabarelli} the Schroedinger equation for the
propagation of neutrinos in matter and assuming the Earth density
profile of Ref.~\cite{PREM}. For all currently allowed values of
$\Delta m^2_{21}/|\Delta m^2_{32}|$, three family interference effects
are negligible in the multi-GeV neutrino sample~\cite{Petcov:2005rv}
($E>1.5$~GeV) and, therefore, the observables are completely CP-blind.

\section{Experimental setup}
\label{sec:setup}

The experimental setup considered in this paper closely follows the
configuration studied in Ref.~\cite{Donini:2006tt}. It exploits a
possible novel source operating in $\nu_e \rightarrow \nu_\mu$ mode,
i.e. a Beta Beam leveraging an upgraded injection complex for the LHC.
The Beta Beams are pure sources of electron-neutrinos obtained
producing, accelerating and stacking beta-unstable isotopes. The most
advanced machine study regarding this novel approach is currently in
progress within the Eurisol Design Study group~\cite{eurisol}; it is based on
existing CERN accelerating machines (PS and SPS) and uses $\helio$ and
$\neon$ as $\bar{\nu}_e$  and $\nu_e$ emitters, respectively.

\begin{figure}[htb]
\centering
\epsfig{file=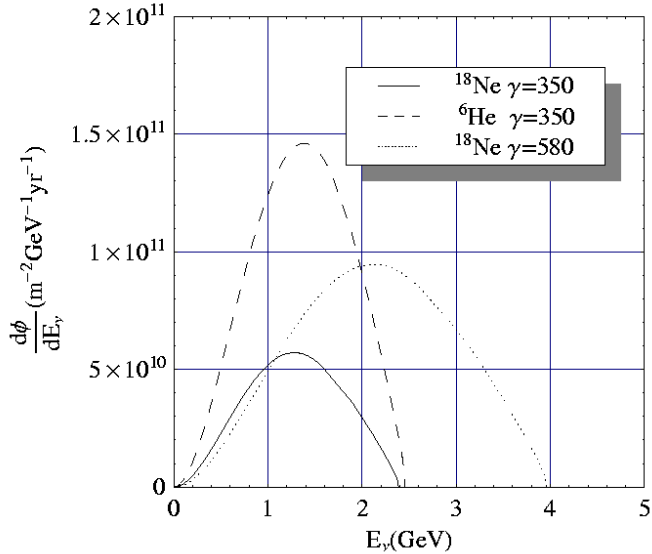,width=9cm}
\caption{Neutrino spectra at the far location.}
\label{fig:spectra}
\end{figure}

Compared with such baseline configuration, the setup of
\cite{Donini:2006tt}, which was originally proposed in
Ref.~\cite{highebb} for a massive water Cherenkov detector, considers
a higher energy proton injector to the LHC (the ``Super-SPS'' instead
of the existing SPS) employed to increase the $\gamma$ of the ions; in
this case, the neutrinos have energies exceeding 1~GeV. The
corresponding spectra are recalled in Fig.~\ref{fig:spectra}. They are
computed assuming the far detector to be located 730~km from CERN
(CERN-to-Gran Sasso distance) and the ions accelerated up the same
$\gamma$ ($\gamma = 350$) both for $\helio$ and for $\neon$
(``$\gamma=350,350$ option''\footnote{$\helio$ and $\neon$ having
nearly the same Q-value, equal $\gamma$ corresponds to equal average
neutrino energy.}) or up to the maximum rigidity allowed by the
Super-SPS ($\gamma=580$ for $\neon$, $\gamma=350$ for $\helio$).  At
these energies, final state muons produced by $\nu_\mu$
charged-current interactions have a range significantly larger than
the pion interaction length in iron. Hence, muon identification
against the bulk of $\nu_e$ CC and NC interactions becomes possible
even for high density iron detectors. Differently from the case of the
Neutrino Factories, the far detector does not require charge
identification capability: the initial state $\nu_e$ beam is free from
$\bar{\nu}_\mu$ contaminations and $\mu^+$ production from charm or
tau decays is kinematically suppressed. In this context, the only
advantage of magnetization comes from the additional suppression of
the NC background with a pion misidentified as a muon.
 
In Ref.~\cite{Donini:2006tt}, the detector was based on glass
Resistive Plate Chambers interleaved with 4~cm thick vertical iron
slabs.  The fiducial mass is 40~kton i.e. the detector can be
accommodated in one of the existing LNGS experimental halls.  The RPC
are housed in a 2~cm gap and the signal is readout on external pick-up
electrodes segmented in $2 \times 2$ ~cm$^2$ pads. This detector can
be divided into large modules (about 15~m length) and magnetized by
copper coils running through the slabs.  This design, which resembles
the toroidal configuration of MINOS, was studied in detail by the
MONOLITH Collaboration in 1999~\cite{monolith_proposal} and it has
been recently revived by the Indian proposal
INO~\cite{ino,Gandhi:2004bj}. Hence, the far detector of
\cite{Donini:2006tt} represents a unique atmospheric neutrino
detector. In particular, MONOLITH studies\footnote{The interest for a
vertical slab configuration was originally motivated by the possible
exploitation of MONOLITH at CNGS (see Ref.\cite{monolith_proposal}
Chap.2).} have demonstrated that the vertical orientation of the steel
does not compromise the capability of reconstructing the oscillation
pattern of atmospheric neutrinos.  Moreover, the atmospheric analysis
profits of the higher granularity needed for the accelerator-based
$\nu_\mu$ appearance search.

In the following, we consider a 40 kton magnetized detector running at
a Beta Beam for 10 years. The accelerator facility is assumed to
provide $2.9{\times} 10^{18}~\helio$ and $1.1{\times} 10^{18}~\neon$
decays per year. As in Ref.~\cite{Donini:2006tt}, we stress that no
solid estimate of fluxes are available for the Super-SPS option: we
refer to \cite{Donini:2006tt} for a systematic study of the
sensitivity as a function of fluxes.

\section{Analysis of atmospheric data}
\label{sec:analysis}

The simulation of atmospheric events at the far detector for the Beta
Beam is similar to the one recently implemented by
MINOS~\cite{minos_atm}: flux expectations are based on the
Bartol~96~\cite{bartol96} model and interactions have been
calculated with GRV94~\cite{grv94} parton distributions including
contributions from quasi-elastic and single pion
production~\cite{tabarelli}. The detector response has been fully
simulated with GEANT3~\cite{geant} assuming the magnetic field
maps resulting from the coil arrangement of MONOLITH. The average
magnetic field along the slabs is $\sim$1.3~T.

As discussed in Sec.~\ref{sec:prob}, a substantial enhancement of the
sensitivity to hierarchy can be obtained tuning the event selection to
identify the region where resonant enhancement occurs.  The resonance
appears as a perturbation to the original $L/E$ pattern for a given
earth matter density. For $\cos \theta>0.85$ the neutrinos cross the
earth core and resonant conversions can appear already at few GeV
($E_R \simeq 3$~GeV). At smaller $\cos \theta$ $\nu$'s intersects the
Earth mantle and crust and resonances appear at larger $E_R$ ($\sim 7$
and $\sim 11$~GeV respectively). Hence, a two dimensional analysis
either in $E$ versus $L$ or, equivalently, in $L/E$ versus $cos
\theta$ is appropriate to extract information on the existence and
location of Earth matter resonances.  The choice of $L/E$ a variable
allows the implementation of the Monolith criteria for the
observability of the sinusoidal pattern~\cite{monolith_proposal} to
the search for resonant conversion.  This approach has been developed
in~\cite{tabarelli}, where the expected resolution on $L$ and $E$ is
computed on an event-by-event basis; here, events are retained only if
the expected resolution is adequate (FWHM on $L/E$ smaller than 50\%)
in the region of interest for matter effect perturbations. A
preselection is applied requiring a minimum muon energy of
1.5~GeV; only internal events are considered: they can be either fully
contained or with a single outgoing track ranging out of the detector
with a visible path-length greater than 4~m.  On average, these
requirements result in a selection efficiency that is negligible below
3~GeV and nearly constant ($\sim$55\%) above 5~GeV.  At these
energies, charge misidentification is well below 5\% and therefore,
rather pure $\mu^+$ and $\mu^-$ samples can be collected.

\begin{figure}[htb]
\centering
\epsfig{file=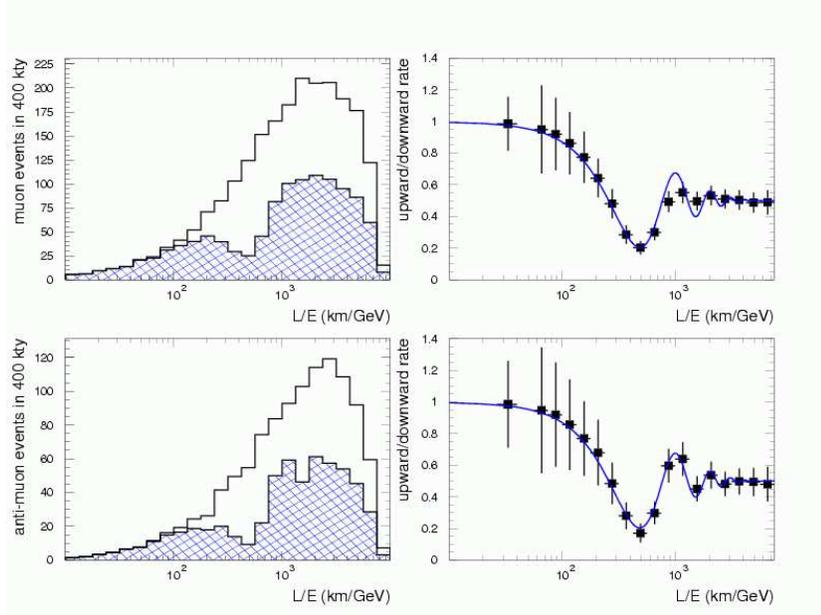,width=11cm}
\caption{L/E distribution of selected events corresponding to
400~kton$\cdot$year (10~y of data taking) for $\Delta m^2_{31}=2.5
\cdot 10^{-3} \ \mbox{eV}^2$, $\sin^2 2\theta_{13}=0.1$ and
$\theta_{23}=45^\circ$ (left upper: neutrinos, left lower:
antineutrinos). The shaded (unshaded) histogram corresponds to the
sample of up-going (down-going) neutrinos. For illustration of the
effect, the ratio upgoing/downgoing is also shown (right) together
with a fit that assumes $\theta_{13}=0^\circ$.  }
\label{fig:LE}
\end{figure}

Fig.~\ref{fig:LE} (left) shows the L/E distribution of selected events
corresponding to 400 kton$\cdot$year (10 y of data taking) for $\Delta
m^2_{31}=2.5 \cdot \ 10^{-3} \ \mbox{eV}^2$, $\sin^2 2\theta_{13}=0.1$
and $\theta_{23}=45^\circ$. In this figure, also down-going events are
shown: they are used as an unoscillated reference sample to reduce the
systematics due to the uncertainty on the initial fluxes.
%(see Fig.~\ref{fig:schema}).
The ratio upgoing/downgoing is shown in Fig.~\ref{fig:LE} (right);
the fit assumes $\theta_{13}=0^\circ$: perturbations due to matter effects
and finite $\theta_{13}$ are visible only for neutrinos since normal
hierarchy is assumed; moreover, distortion with respect to pure
$\nu_\mu \rightarrow \nu_\tau$ oscillations are present mainly beyond
the first minimum of the survival probability.

\section{Sensitivity to mass hierarchy}
\label{sec:sensitivity}

The determination of mass hierarchy from atmospheric neutrinos is
plagued by the uncertainty on the mixing parameters, particularly on
$\theta_{13}$.  Accelerator neutrino experiments (a Beta Beam in the
present case) can provide very strong constraints on this angle, once
marginalized over $\delta$. In particular, the manifold of the allowed
$\theta_{13},\delta$ values is highly reduced by running both in
$\nu_e \rightarrow \nu_\mu$ and in its CP-conjugate mode
(anti-neutrino run). Hence the combination of accelerator and
atmospheric data can be rather effective~\cite{Huber:2005ep,campagne}.

\subsection{Likelihood analysis}

\begin{figure}
\centering
\epsfig{file=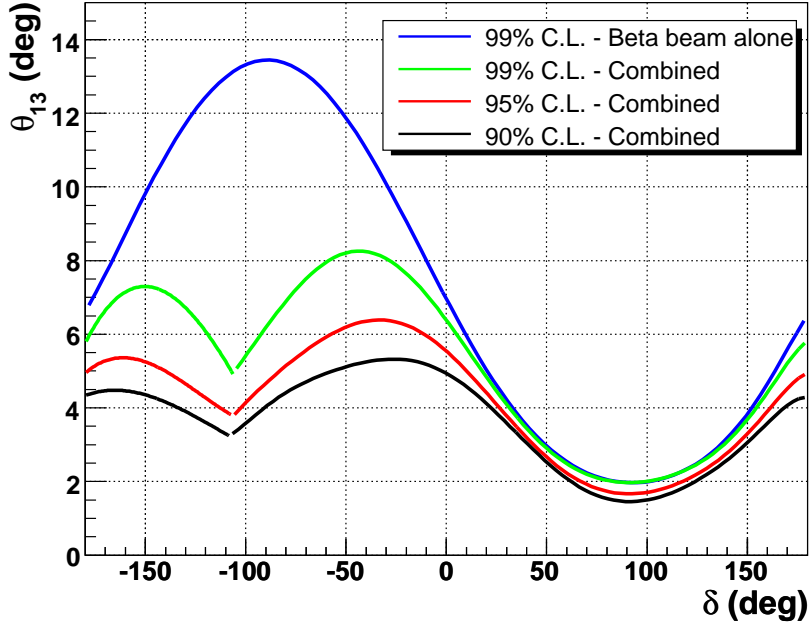,width=12cm}
\caption{Region of the true value of the parameters $\theta_{13}$ and
$\delta$ where the correct neutrino hierarchy (assumed to be normal in
this figure) can be distinguished from the wrong one at 90, 95 and
99\% C.L. Data from the Beta Beam only are also shown.}
\label{fig:sensitivity}
\end{figure}

\begin{figure}
\centering
\epsfig{file=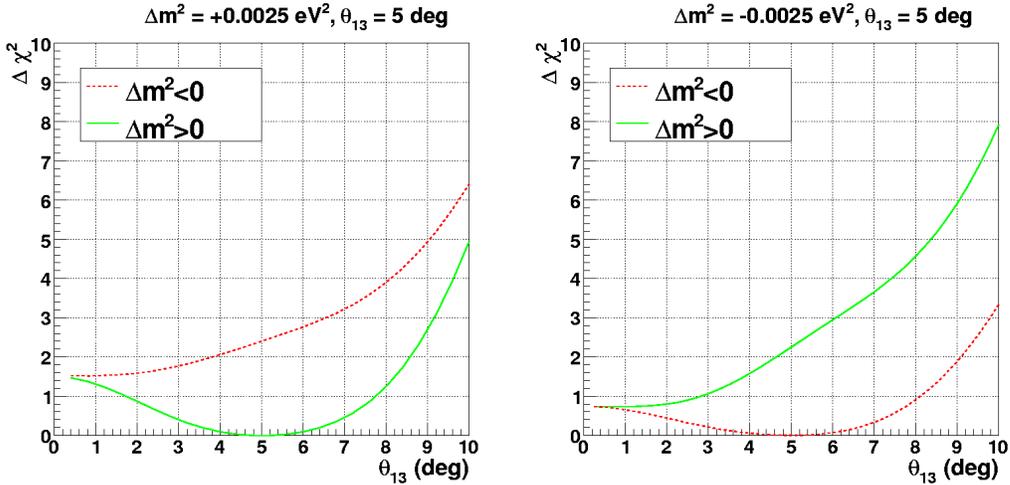,width=14cm}
\caption{Likelihoods of the atmospheric sample ($\Delta \chi^2 \equiv
  -2 Log (L/L_{max})$) as a function of $\theta_{13}$ assuming normal
  hierarchy (left plot) or inverted hierarchy (right plot) as true
  hierarchy and $\theta_{13}=5^\circ$. The continuous green line is
  the likelihood for $\Delta m^2>0$ (right hypotesis in the left plot,
  wrong hypothesis in the right plot) and the dashed red line is for
  $\Delta m^2<0$.}
\label{fig:atmlik}
\end{figure}

The analysis of accelerator data employed in the following does not
differ from the one described in \cite{Donini:2006tt}: we refer to
that reference for the details of event selection and detector
performance. The sensitivity to the sign of $\Delta m^2_{31}$
is computed from a likelihood analysis that combines the allowed $\theta_{13}$
range resulting from the Beta Beam run with the atmospheric likelihood:

\begin{equation}
\label{like} 
\ln {\cal L} = \ln \prod_{i,q}
    \left[ 
    \frac{e^{-A_q\mu_{i,q}} (A_q\mu_{i,q})^{U_{i,q}}}{U_{i,q}!}
    \right]
    - \sum_q \frac{1}{2}\frac{(A_q-1)^2}{\sigma_A^2}
\label{lik_atm}
\end{equation}
where the subscripts $i$ and $q$ are referred to bins in the
($\cos\theta$, $L/E$) plane and of muon charge respectively; $U_{i,q}$
and $A_q \mu_{i,q}$ are the observed and expected number of up-going
neutrino events in the ${i,q}$-th bin and the last term in the
likelihood accounts for the overall normalization 
uncertainty for neutrinos and anti-neutrinos separately: further
details on the treatment of systematics and their relevance for the
mass hierarchy determination are provided below
(Sec.~\ref{sec:systematics}).  Additional penalty terms of the form
$(P-P_{best})^2/2\sigma^2$ are added to (\ref{lik_atm}) to account
for the uncertainty on $P=|\Delta m^2_{31}|$ and
$P=\theta_{23}$. Here, $P_{best}$ is the current best estimate of the
parameter and $\sigma$ the estimated precision. The latter is the
current precision for $\theta_{23}$ and the expected one at the
end of T2K for $|\Delta m^2_{31}|$. For each ``true'' value of
$\theta_{13}$, $\delta$ and the sign of $\Delta m^2_{31}$
%\footnote{We assume $\Delta m^2_{31}$... {\bf DA FARE}}
(e.g. +1 for normal hierarchy), we compute the C.L. of the best fit
for the wrong hypothesis (e.g. sign$(\Delta
m^2_{31})=-1$). Fig.~\ref{fig:sensitivity} shows the region where the
wrong hypothesis is excluded at 90, 95 and 99\% C.L. This figure
demonstrates what stated qualitatively in Sec.~\ref{sec:prob}: due to
the insensitivity to the Dirac phase, the cancellation effect that
equals the rates of $\mu^+$ and $\mu^-$ at the Beta Beam does not take
place with atmospherics. Hence, atmospheric data are relevant and,
actually, dominate the sensitivity to mass hierarchy in most of the
Beta Beam blind region. In the present scenario, hierarchy can be
determined at 90\% C.L. down to $\theta_{13} \simeq 4^\circ$. At
positive (negative) $\delta$ and normal (inverted) hierarchy, the Beta
Beam data dominates and sign($\Delta m^2_{31}$) can be extracted down
to $\theta_{13} \simeq 2^\circ$.  Compared with stand-alone analysis
of atmospheric neutrinos with high density detectors~\cite{tabarelli},
where some sensitivity can be achieved only for $\theta_{13}$ values
very close to current CHOOZ limits, here the accelerator-based
determination of this parameter pushes the sensitivity to the mass
hierarchy down to $\sin^2 2\theta_{13} \simeq 0.02$ even for very
unfavorable values of the Dirac phase $\delta$.  This is illustrated
in Fig.~\ref{fig:atmlik} where the likelihoods for the atmospheric
sample only are shown for $\theta_{13}=5^\circ$.  The
left (right) plots assume normal (inverted) hierarchy as the true
hierarchy. In general, Fig.~\ref{fig:atmlik} shows that hierarchy
separation cannot be achieved if $\theta_{13}$ is completely
unconstrained: on the other hand, even for the most unfavorable values
of $\delta$, the Beta Beam provides stringent limits to $\theta_{13}$
that makes the atmospheric sample a sensitive tool for hierarchy
determination. If the true hierarchy is inverted, distortions will
appear in the antineutrino sample, i.e. in the atmospheric sample with
lower statistics due to the difference between $\nu_\mu$ and
$\bar{\nu}_\mu$ cross sections. However, the discrimination power
depends on the difference between the higher statistics $\nu_\mu$
sample and the lower statistics $\bar{\nu}_\mu$ one and, as visible
from Fig.~\ref{fig:atmlik}, the power of the test is practically the
same both assuming normal and inverted ``true'' hierarchy.

The use of the Monolith criteria for event selection improves
substantially the sensitivity with respect to analyses where flat
resolutions have been assumed~\cite{Petcov:2005rv}.  In spite of their
limited mass, high density atmospheric detectors compete with
megaton-size water Cherenkov's thanks to the capability of measuring
the sign of the multi-GeV sample. On the other hand, due to larger
statistics, accelerator data are sensitive to perturbations induced by
the sign of $\Delta m^2$ up to $\sin^2 2\theta_{13} \simeq 0.03-0.02$
(see Fig.16 of Ref.\cite{campagne}). Finally, it is worth noting that
significantly better results (sensitivities to the hierarchy down to
$\sin^2 2\theta_{13} < 0.01$) could be achieved by dedicated detectors
located at the magic baseline without the use of atmospheric
sample~\cite{agarwalla}. These setups, however, requires either large
boosters for the beta-beam ions (LHC) or novel ions with larger
Q-values and increased flux to compensate for the losses at the far
location~\cite{litium}. Moreover, no information on CP violation in
the leptonic sector would be accessible and a second detector at a
closer distance from the source would be needed.

\subsection{Systematic uncertainties for the atmospheric sample}
\label{sec:systematics}

Systematic uncertainties on the rate of multi-GeV and sub-GeV samples
of atmospheric neutrinos have been studied in depth in literature due
to their relevance for the determination of the leading oscillation
parameters $\Delta m_{13}$ and $\sin^2 \theta_{23}$. Moreover, they
affect the pattern of muon disappearance and the perturbation of the
$\nu_e$ sample and, therefore, they can be limiting factors in the
determination of the mass hierarchy and $\theta_{13}$. In our case,
the knowledge of $\theta_{13}$ is completely dominated by accelerator
data while the determination of the mass hierarchy is mainly related
to the location of the MSW resonance in the neutrino or antineutrino
sample. Hence, several systematic contributions that currently limit
the knowledge of the leading parameters at atmospheric detectors
(overall flux and cross section normalization, $\nu_\mu / \nu_e$
ratio, zenith and energy dependence of the differential fluxes) play
here a minor role. In order to probe quantitatively such statement, in
the present analysis systematics uncertainties have been treated
following the approach developed by the Monolith collaboration and
based on up-down differential
normalization~\cite{monolith_proposal,picchi}. For each bin in the
$L/E$ distribution of the up-going neutrinos, the unoscillated sample
is taken from the corresponding down-going. Here ``up-going'' means
that neutrinos reach the detector coming from below the horizon ($\cos
\theta>0$); the ``corresponding down-going'' are neutrinos whose Nadir
angle $\theta'$ is such that $\theta'=\pi-\theta$. This subsample
provides a nearly theory-free normalization since the uncertainties on
the flux and cross sections affecting the population of such bin
cancel out because of the spherical symmetry and the isotropy of the
primary cosmic ray flux. Further details and interesting subtleties
can be found in the Monolith proposal~\cite{monolith_proposal}. For
instance, geomagnetic effects can introduce an asymmetry in the
up/down sample that could be interpreted as a perturbation of the
$L/E$ pattern.  However, for the multi-GeV sample selected in this
analysis, with a minimum neutrino energy of 3 GeV (see
Sec.~\ref{sec:analysis}), this effect is
negligible~\cite{Agrawal:1995gk}.
Yet, an overall penalty term in the flux normalization has been
introduced (see Eq.~\ref{lik_atm}), to cope with the $\nu_\mu/\nu_e$
ratio uncertainty, as the proposed detector cannot directly measure
the $\nu_e$ flux.  This approach is somewhat conservative: in most of
the cases, the present theoretical knowledge of the atmospheric fluxes
exceed the statistical precision of the down-going samples for bin
widths comparable to the detector resolution. However, for the present
case where systematics play a less important role than for the leading
parameter analysis, this technique simplifies remarkably the treatment
of systematics without compromising the sensitivity to the mass
hierarchy. To ease comparison with other analyses (See e.g. Table 1
of~\cite{Petcov:2005rv}) we note that the statistical uncertainty of
the down-going sample in the region where the oscillation dip at
$\theta_{13}$ is located is of about 9\%, while typical differential
uncertainties in the rate of multi-GeV $\nu_\mu$ are of the order of
5\%. Moreover, the down-going normalization
sample is measured for neutrinos and antineutrinos separately, hence
assuming that the uncertainty on the $\nu_\mu$/$\bar{\nu}_\mu$ ratio
could also have a dependence in energy and in direction, while in most
of the analyses a flat uncertainty of about 5\% is considered. As a
consistency check, we artificially decreased the statistical
uncertainty of the down-going sample up to 3\% noting that the
sensitivity to the sign of $\Delta m^2_{31}$ remains practically
unchanged.  This is in agreement to what already observed by some
authors, i.e. that ``the discrimination between normal and inverted
hierarchy is based on a very characteristic signal consisting of
pronounced structures in the $E$ and $\cos \theta$ distributions,
which cannot easily mimicked by the systematic
effects''~\cite{Petcov:2005rv}.

The most relevant systematic uncertainty of instrumental origin is
related to the pollution of the neutrino sample with antineutrinos due
to charge misassignment. In most of the analyses presented in
literature, an overall uncertainty on the knowledge of the charge
(typically of 5\%) is assumed independently of the topology of event.
The present analysis, instead, is based on a full GEANT3 simulation of
the detector, followed by a reconstruction stage with track
fitting. Neutrino flavour assignment is based on the reconstructed
muon charge, thus encompassing a realistic energy and event shape
dependency, as far as GEANT3 can predict the behaviour of magnetised
iron calorimeters\footnote{Solid testbeds are based, e.g., on the CDHS
and MINOS data, on non-compensated hadronic calorimeters at colliders
and on Monolith testbeams~\cite{babymonolith}.}.
The charge misidentification
systematics is the most relevant detector-dependent systematics in the
atmospheric analysis, while it is immaterial in the Beta Beam
analysis, where $\pi$/$\mu$ misidentification dominates. This is the
reason why we assumed the systematics of the accelerator and
atmospheric data fully uncorrelated.

\section{Conclusions}
\label{conclusions}

Detailed atmospheric neutrino studies can be done in several detectors
that have been built or proposed for long-baseline accelerator
facilities. In particular, high density magnetized detectors can study
multi-GeV atmospheric neutrinos and distinguish between $\nu_\mu$ and
$\bar{\nu}_\mu$. These data can be combined in a highly non trivial
manner to improve the sensitivity to the neutrino mass hierarchy.  In
this paper we considered in particular a high energy Beta Beam with a
40~kton iron calorimeter located at the CERN to Gran Sasso distance.
The atmospheric data collected by this detector allows the
determination of the neutrino hierarchy down to $\theta_{13} \simeq
4^\circ$ in the region of $\delta$ where the corresponding Beta Beam
data cannot constrain the sign of $\Delta m^2_{31}$. Far from this
blind region, the Beta Beam data dominates and sign($\Delta m^2_{31}$)
can be extracted down to $\theta_{13} \simeq 2^\circ$ at 90\% C.L.

\section*{Acknowledgments}
 
We wish to express our gratitude to S.~Petcov, S.~Agarwalla and
A.~Blondel for several useful discussions during the WIN07 workshop in
Kolkata.

%%%%%%%%%%%%%%%%%%%%%%%%%%%%%%%%%%%%%%%%%%%%%%%%%%%%%%%%%%%%%%%%%%%%%%%%%%
%BIBLIOGRAPHY
%%%%%%%%%%%%%%%%%%%%%%%%%%%%%%%%%%%%%%%%%%%%%%%%%%%%%%%%%%%%%%%%%%%%%%%%%%


\begin{thebibliography}{999}
\bibitem{pdg}   W.~M.~Yao {\it et al.}  [Particle Data Group],
  J.\ Phys.\ G {\bf 33} (2006) 1 and references therein.
\bibitem{evidence} Q.R.~Ahmad {\it et al.}  [SNO Coll.], Phys.\ Rev.\
Lett.\ {\bf 87} (2001) 071301; B.~Aharmim {\it et al.}  [SNO
Collaboration], Phys.\ Rev.\ C {\bf 72} (2005) 055502; T.~Araki {\it
et al.}  [KamLAND Collaboration] Phys.\ Rev.\ Lett.\ {\bf 94} (2005)
081801; Y.~Ashie {\it et al.}  [Super-Kamiokande Collaboration],
Phys.\ Rev.\ D {\bf 71} (2005) 112005; M.~Ambrosio {\it et al.}
[MACRO Collaboration], Phys.\ Lett.\ B {\bf 566} (2003) 35; .~Aliu
{\it et al.}  [K2K Collaboration], Phys.\ Rev.\ Lett.\ {\bf 94} (2005)
081802.  K.~Abe {\it et al.}  [Super-Kamiokande Collaboration], Phys.\
Rev.\ Lett.\ {\bf 97} (2006) 171801; J.~Hosaka {\it et al.}
[Super-Kamiokande Collaboration], Phys.\ Rev.\ D {\bf 74} (2006)
032002; D.~G.~Michael {\it et al.}  [MINOS Collaboration],
  Phys.\ Rev.\ Lett.\  {\bf 97} (2006) 191801.
\bibitem{opera_NJP} R.~Acquafredda {\it et al.}  [OPERA
  Collaboration], New J.\ Phys.\ {\bf 8} (2006) 303.
\bibitem{reactor} M.~Apollonio {\it et al.}  [CHOOZ Collaboration],
  Eur.\ Phys.\ J.\ C {\bf 27} (2003) 331; M.~Apollonio {\it et al.}
  [CHOOZ Collaboration], Phys.\ Lett.\ B {\bf 466} (1999) 415;
  F.~Boehm {\it et al.} [Palo Verde Collaboration], Phys.\ Rev.\ D {\bf 64}
  (2001) 112001.
\bibitem{review} For a review see A.~Guglielmi, M.~Mezzetto,
P.~Migliozzi and F.~Terranova, arXiv:hep-ph/0508034 (in D.~Bettoni 
{\it et al.},  Phys. Rep. {\bf 434} (2006) 47, pp.68-87).
\bibitem{nufact} S.~Geer, Phys.\ Rev.\ D {\bf 57} (1998)
6989 [Erratum-ibid.\ D {\bf 59} (1999) 039903]; A.~De Rujula,
M.~B.~Gavela and P.~Hernandez, Nucl.\ Phys.\ B {\bf 547} (1999) 21.
\bibitem{betabeam}
P.~Zucchelli, Phys.\ Lett.\ B {\bf 532} (2002) 166.
\bibitem{waterch} J.~Bernabeu, S.~Palomares Ruiz and S.~T.~Petcov,
  Nucl.\ Phys.\ B {\bf 669} (2003) 255; M.~C.~Gonzalez-Garcia and
  M.~Maltoni, Eur.\ Phys.\ J.\ C {\bf 26} (2003) 417;
  M.~C.~Gonzalez-Garcia, M.~Maltoni and A.~Y.~Smirnov, Phys.\ Rev.\ D
  {\bf 70} (2004) 093005
\bibitem{Huber:2005ep} P.~Huber, M.~Maltoni and T.~Schwetz, Phys.\
Rev.\ D {\bf 71} (2005) 053006.
\bibitem{campagne} J.~E.~Campagne, M.~Maltoni,
M.~Mezzetto and T.~Schwetz, JHEP {\bf 0704} (2007) 003.
\bibitem{highebb} J.~Burguet-Castell, D.~Casper, J.~J.~Gomez-Cadenas,
P.~Hernandez and F.~Sanchez, Nucl.\ Phys.\ B {\bf 695} (2004) 217;
J.~Burguet-Castell, D.~Casper, E.~Couce, J.~J.~Gomez-Cadenas and
P.~Hernandez, Nucl.\ Phys.\ B {\bf 725} (2005) 306.
\bibitem{Donini:2006tt} A.~Donini, E.~Fernandez-Martinez,
P.~Migliozzi, S.~Rigolin, L.~Scotto Lavina, T.~Tabarelli de Fatis and
F.~Terranova, Eur.\ Phys.\ J.\ C {\bf 48} (2006) 787
\bibitem{highebb2} P.~Huber, M.~Lindner, M.~Rolinec and W.~Winter,
Phys.\ Rev.\ D {\bf 73} (2006) 053002;
F.~Terranova, A.~Marotta, P.~Migliozzi and M.~Spinetti, Eur.\ Phys.\
J.\ C {\bf 38} (2004) 69.
\bibitem{agarwalla}  S.~K.~Agarwalla,
A.~Raychaudhuri and A.~Samanta, Phys.\ Lett.\ B {\bf 629} (2005) 33;
S.~K.~Agarwalla, S.~Choubey and A.~Raychaudhuri,
  Nucl.\ Phys.\  B {\bf 771} (2007) 1.
\bibitem{litium}C.~Rubbia, A.~Ferrari, Y.~Kadi and V.~Vlachoudis,
Nucl.\ Instrum.\ Meth.\ A {\bf 568} (2006) 475; A.~Donini and
E.~Fernandez-Martinez, Phys.\ Lett.\ B {\bf 641} (2006) 432.
\bibitem{Petcov:2005rv}
  S.~T.~Petcov and T.~Schwetz, Nucl.\ Phys.\  B {\bf 740} (2006) 1
\bibitem{chen} C.~H.~Albright and M.~C.~Chen,
  Phys.\ Rev.\  D {\bf 74} (2006) 113006.
\bibitem{disappearance} A. de Gouvea, J. Jenkins and B. Kayser,
Phys. Rev. D {\bf 71} (2005) 113009; H. Minakata, H. Nunokawa,
S. J. Parke, R. Zukanovich Funchal, Phys. Rev. D {\bf 74} (2006)
053008; H.~Minakata, H.~Nunokawa, S.~J.~Parke and R.~Z.~Funchal,
Phys.\ Rev.\ D {\bf 76} (2007) 053004 [Erratum-ibid.\ D {\bf  76}
(2007) 079901].
\bibitem{cervera}
A.~Cervera et~al., Nucl. Phys. B {\bf 579}~(2000)~17, erratum ibid.
Nucl. Phys. B {\bf 593}~(2001)~731.
\bibitem{freund}
M.~Freund, Phys. Rev. D {\bf 64}~(2001)~053003.
\bibitem{supersps} O.~Bruning et al., ``LHC luminosity and energy
upgrade: A feasibility study,'' CERN-LHC-PROJECT-REPORT-626, 2002;
W.~Scandale, Nucl.\ Phys.\ Proc.\ Suppl.\ {\bf 154} (2006) 101.
\bibitem{nova} D.~S.~Ayres {\it et al.}  [NOvA Collaboration],
  arXiv:hep-ex/0503053.
\bibitem{magic}
  V.~Barger, D.~Marfatia and K.~Whisnant,
  Phys.\ Rev.\  D {\bf 65} (2002) 073023; P.~Huber and W.~Winter,
  Phys.\ Rev.\  D {\bf 68} (2003) 037301.
\bibitem{Smirnov:2006sm} A.~Y.~Smirnov, arXiv:hep-ph/0610198.
\bibitem{tabarelli} T.~Tabarelli de Fatis,
  Eur.\ Phys.\ J.\  C {\bf 24} (2002) 43.
\bibitem{PREM}  A.M.~Dziewonski and D.L.~Anderson, 
                Phys. Earth Planet. Inter., {\bf 25} (1981) 297
\bibitem{eurisol} M.~Benedikt, A.~Fabich, S.~Hancock, M.~Lindroos,  
EURISOL DS/TASK12/TN-05-03. 
\bibitem{monolith_proposal} N.Y. Agafonova {\it et al.} [MONOLITH
Collaboration], LNGS-P26-2000, LNGS-P26-00, CERN-SPSC-2000-031,
CERN-SPSC-M-657.
\bibitem{ino} 
M.~S.~Athar {\it et al.}  [INO Collaboration],
``India-based Neutrino Observatory: Project Report. Vol.I,'', INO-2006-01.
\bibitem{Gandhi:2004bj}
R.~Gandhi, P.~Ghoshal, S.~Goswami, P.~Mehta and S.~Uma Sankar,
Phys.\ Rev.\  D {\bf 73}  (2006) 053001.
\bibitem{minos_atm} P.~Adamson {\it et al.} [MINOS Collaboration],
Phys.\ Rev.\ D {\bf  75} (2007) 092003. 
\bibitem{bartol96} V.~Agrawal, T.K.~Gaisser, P.~Lipari and T.~Stanev,
 Phys. Rev. D {\bf  53} (1996) 1314.
\bibitem{grv94} M.~Gluck, E.~Reya and A.~Vogt,
                Z. Phys. C {\bf  67} (1995) 433.
\bibitem{geant} GEANT - Detector Description and Simulation Tool CERN
Program Library Long Writeup W5013.
\bibitem{picchi} P.~Picchi and F.~Pietropaolo, ICGF Internal Note 344/1997,
available as CERN Preprint SCAN, SCAN-9710037.
\bibitem{Agrawal:1995gk}
  V.~Agrawal, T.~K.~Gaisser, P.~Lipari and T.~Stanev,
  Phys.\ Rev.\  D {\bf 53} (1996) 1314.
\bibitem{babymonolith} M.~Ambrosio {\it et al.}, Nucl.\ Instrum.\
  Meth.\ A {\bf 456} (2000) 67.

\end{thebibliography}
\end{document}